\documentclass[prc,amsmath,twocolumn,showkeys,superscriptaddress]{revtex4}

\usepackage{gensymb}
\usepackage{multirow}
\usepackage{graphicx,color}
\usepackage{amssymb}
\usepackage{enumerate}
\usepackage{verbatim}
\usepackage{natbib}
\usepackage{dsfont}
\usepackage{float}
\usepackage[outdir=./]{epstopdf}
\usepackage{ulem}

\begin{document}
  \newcommand {\nc} {\newcommand}
   \nc {\etal} {\textit{et al.}}
  \nc {\Sec} [1] {Sec.~\ref{#1}}
  \nc {\IBL} [1] {\textcolor{black}{#1}} 
  \nc {\IR} [1] {\textcolor{red}{#1}} 
  \nc {\IB} [1] {\textcolor{blue}{#1}} 
  \nc {\IG} [1] {\textcolor{green}{#1}}
  \nc {\Sch} {Schr\"odinger }
  \nc {\beq} {\begin{eqnarray}}
  \nc {\eeq} {\nonumber \end{eqnarray}}
  \nc {\eeqn}[1] {\label {#1} \end{eqnarray}}
  \nc {\ve} [1] {\mbox{\boldmath $#1$}}
  \nc{\pderiv}[2]{\cfrac{\partial #1}{\partial #2}}  
  \nc {\flim} [2] {\mathop{\longrightarrow}\limits_{{#1}\rightarrow{#2}}}
  \nc{\deriv}[2]{\cfrac{d#1}{d#2}}

\title{Considering non-locality in the optical potentials within eikonal models}

\author{C. Hebborn}
\email{hebborn@frib.msu.edu}
\affiliation{Facility for Rare Isotope Beams, East Lansing, MI 48824}
\affiliation{Lawrence Livermore National Laboratory, P.O. Box 808, L-414, Livermore, California 94551, USA}
\author{F.~M.~Nunes}
\email{nunes@nscl.msu.edu}
\affiliation{National Superconducting Cyclotron Laboratory, Michigan State University, East Lansing, MI 48824}
\affiliation{Department of Physics and Astronomy, Michigan State University, East Lansing, MI 48824-1321}

\date{\today}


\begin{abstract}
\begin{description}
\item[Background:] For its simplicity, the eikonal method is the tool of choice  to analyze nuclear reactions at high energies ($E>100$~MeV/nucleon), including knockout reactions. However, so far, the effective interactions used in this  method are assumed to be fully local.  
\item[Purpose:] Given the recent studies on non-local optical potentials, in this work we assess whether non-locality in the optical potentials is expected to impact reactions at high energies and then explore different avenues for extending the eikonal method to include non-local interactions.
\item[Method:] We compare angular distributions obtained  for  non-local interactions (using the exact R-matrix approach for elastic scattering and the adiabatic distorted wave approximation for transfer) with those obtained using their local-equivalent interactions. 
\item[Results:]  Our results show that transfer observables are significantly impacted by non-locality in the high-energy regime. Because knockout reactions are dominated by stripping (transfer to inelastic channels), non-locality is expected to have a large effect on knockout observables too. Three approaches are explored for extending the eikonal method to non-local interactions, including an iterative method and  a perturbation theory. 
\item[Conclusions:]   None of the derived extensions of the eikonal model provide a good description of elastic scattering. This work suggests that non-locality removes the formal simplicity associated with the eikonal  model.
\end{description}
\end{abstract}


\maketitle

\section{Introduction}
\label{intro}

The most versatile probe into the structure of matter in its extreme forms are nuclear reactions. Experimental programs around the world have used a wide variety of reactions  to extract a diverse range of information on the properties of different isotopes (e.g. \cite{prl-yang2021,prc-wang2021,prc-kay2021} for some of the most recent examples). In all these cases, the measured cross sections are interpreted through a reaction model. Regardless of the sophistication level of the reaction model used, the effective interactions between the cluster parts are an essential ingredient to all. These effective interactions are known as optical potentials, because they include an imaginary term which accounts for all those processes that are not explicitly included in the model.

Even for the simplest case, the effective interaction between a nucleon and a nucleus, optical potentials are intrinsically non-local, due to antisymmetrization and coupling to excited modes \cite{F58,Canton_prl2005,Fraser_epja2008}. These features emerge naturally in those potentials derived microscopically from \textit{ab initio} many-body approaches (e.g. \cite{rotureau2017,rotureau2018,idini2019,WLH20}). Since the early days when Perey and Buck developed their non-local optical potential, with a Gaussian non-locality \cite{perey1962}, not many groups have invested in a global non-local potential (see \cite{tian2015,lovell2017} for  recent works on the topic). Most of the phenomenological global optical potentials widely used in the field are approximated to be local for numerical convenience~\cite{kd2003,ch89,bg69}.

Over the last few years, a large body of work demonstrates that including non-locality explicitly in the optical potential significantly affects the calculated reactions observables \cite{Titus_prc2014,Ross_prc2015,Timofeyuk_prl2013,Waldecker_prc2016,Titus_PRC2016,Ross_prc2016,prc-li2018}.
These studies have focused primarily on $(d,p)$ transfer reactions in the energy range $E_{lab}< 50$~MeV. The effects of non-locality manifest in the short-range part of the bound and scattering wave functions. These effects are then picked up in amplitudes for transfer, evidently dependent on off-shell behavior.  In many cases, and particularly for heavier nuclei, the resulting transfer angular distributions calculated with non-local optical potentials differ considerably  in shape and in magnitude from the local counterparts, and would inevitably lead to discrepancies in the extracted spectroscopic factors.

Given the increasing interest in experiments at higher beam energy using knockout and breakup reactions, it is important to understand the role of non-locality for energies above 100~MeV/nucleon. These reactions have been extensively used to extract structure information, but their interpretation rely primarily on eikonal methods with local interactions~\cite{G59,HM85,HT03}. Until now, there have been no investigations on the effects of non-locality in the optical potential within this energy regime. Neither have the eikonal methods been extended to include the capability of non-locality in the interactions. Incorporating non-local interactions in the eikonal theory is non-trivial and therefore it is advisable to first inspect whether such extensions would be necessary.

There are two main questions  this work is addressing: 1) how large are the effects of non-locality  in the higher energy regime? and 2) what are the challenges in including non-locality explicitly in the eikonal methods to describe reactions? Since the methods for transfer reactions have already been extended to include non-locality explicitly,  in the first part of this work we study transfer reactions at energies that are higher than what would normally be used for this type of reaction ($>100$~MeV/nucleon). Because the stripping mechanism, corresponding to the non-elastic channels where the  nucleon is absorbed by the target,  is the largest  contributor to  knockout cross sections~\cite{LB21}, if effects turn out to be large for transfer, then effects can also be expected to be large for knockout. Once that is established, we consider a couple of different paths to include non-locality explicitly in the eikonal formalism.

This paper is organized as follows: in Sec.~\ref{Sec2} we present a brief overview of the two-body local and non-local scattering problems. Next, we consider results for transfer reactions (in Sec.~\ref{Sec3}) and  elastic scattering (in Sec.~\ref{Sec4}). In Sec.~\ref{Sec5} we discuss non-local extensions of the eikonal model. Conclusions are drawn in Sec.~\ref{Conclusions}.


\section{ Theoretical background}\label{Sec2} 

We review in this section the description of  the elastic-scattering channel and refer the interested reader to Refs.~\cite{JT74,Titus_CPC2016,TitusThesis} for the calculation of transfer observables. 

As usual, we simplify the many-body problem to a two-body one, in which both the projectile $P$ and the target $T$ are assumed structureless and spinless. This reduction comes at a cost, the $P$-$T$ interaction is simulated through an optical potential, which includes an imaginary part   modeling effectively the inelastic channels~\cite{F58,BC12}.

\subsection{Local interaction}
If the $P$-$T$ interaction is simulated by a local central optical potential $V_L$, the two-body system is described by the following \Sch equation~\cite{BC12}
\beq
\left[\frac{P^2}{2\mu}+V_L({\ve{R}})\right]\Psi(\ve{R}) = E\ \Psi(\ve{R}),
\eeqn{eq1}
where $\mu=m_Pm_T/(m_P+m_T)$ is the $P$-$T$ reduced mass, $\ve P$ and $\ve R\equiv(\ve b,Z)$ are respectively the $P$-$T$ relative momentum and coordinate and $E$ is the total energy of the system. This equation is solved with the  condition that initially the projectile  propagates towards the target along the $Z$-axis with a velocity $v=\hbar K/\mu$ and momentum $\hbar K$, i.e.,
\beq
\Psi(\ve{R})\flim{Z}{- \infty}\exp(iKZ).
\eeqn{eq2}

There are several approaches typically used to solve this two-body problem  exactly. In this paper, we obtain the exact solution using the R-matrix method combined with the Lagrange-mesh method~\cite{DB10,B15}. One of the main  advantages of this technique  is its straightforward generalization to non-local potentials.

The eikonal approximation~\cite{G59} reflects the fact   that at high enough energy the $P$-$T$ relative motion does not differ much from the initial plane wave. It hence factorizes this plane wave out of the wave function
\beq
\Psi(\ve{R})=\exp(iKZ)\ \widehat{\Psi}(\ve{R}),
\eeqn{eq3} 
and assumes that   $\widehat{\Psi}$ varies smoothly with $\ve R$. Furthermore, the eikonal approximation neglects  the second-derivatives of $\widehat{\Psi}$, simplifying Eq.~\eqref{eq1} into~\cite{BC12}
\beq
i\hbar v \pderiv{}{Z}\widehat{\Psi}(\ve{b},Z)=V_L({\ve{b}},Z)\ \widehat{\Psi}(\ve{b},Z)
\eeqn{eq4}
where $\ve b$ is the transverse coordinate of $\ve R$. 
This simplified \Sch equation can be solved analytically and its solutions behave asymptotically as
\beq
\widehat{\Psi} (\ve{b},Z)\underset{Z\to+\infty}{\longrightarrow}\exp\left[-\frac{i}{\hbar v}\int_{-\infty}^{+\infty} V_L(\ve{b},Z)\ \mathrm{d}Z\right].
\eeqn{eq5}
In a semiclassical view, the eikonal solutions can be seen as the projectile following a
straight-line trajectory at constant impact parameter $b$ and accumulating a phase through
its reaction process while interacting with the target.

Elastic-scattering observables depend only on the asymptotic behavior of the wave function, i.e., on the phase in Eq.~\eqref{eq5},  and are therefore  efficiently computed within the eikonal approximation. This model is accurate for  reactions at high enough energy~\cite{HT03,OYI03,B05,BCG05,OB10,PDB12,MC19}. However, the eikonal description  is expected to fail when the wave function differs strongly from a plane wave, i.e., at low energy,   small impact parameters and large scattering angles~\cite{AKTB97,HC17,HC18}.

\subsection{Non-local interaction}

In their most general form, the  optical potentials are non-local~\cite{F58}. This non-locality arises from antisymmetrization
of the many-body wave function and  the couplings between the different channels. When a non-local potential $V_{NL}$ is considered, the   \Sch equation reads~\cite{A65}
\beq
\frac{P^2}{2\mu}\Psi(\ve{R}) +\int d\ve R' \,V_{NL}({\ve{R}},\ve R')\Psi(\ve{R}') = E\ \Psi(\ve{R}),
\eeqn{eq6}
in which the interaction term is obtained through an integration of the wave function and the non-local potential. 
We solve this equation with the R-matrix method using the same initial condition~\eqref{eq2} as in the local case. 

In this article, we  first analyze how non-locality affects transfer [Sec.~\ref{Sec3}] and elastic-scattering [Sec.~\ref{Sec4}] observables  at high energies. Then, we  investigate different extensions of the eikonal approximation to the non-local \Sch equation~\eqref{eq6} in Sec.~\ref{Sec5}.


\section{Effects of non-locality for reactions at high energies}\label{Sec3}

\begin{figure}
	\centering
	\includegraphics[width=\linewidth]{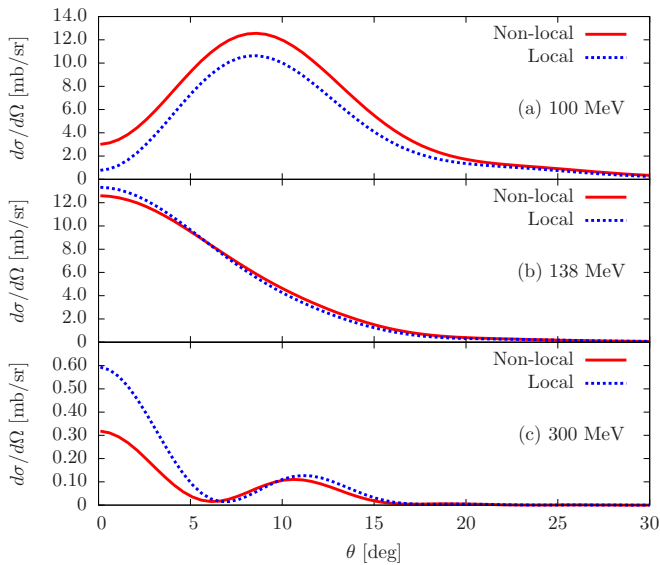}
	\caption{Comparison of the angular distributions for the $\,^{208}{\rm Pb}(d,p)^{209}{\rm Pb}$ at $E_{lab}=100$~MeV (top panel), $E_{lab}=138$~MeV (mid panel)  and $E_{lab}=300$ MeV (bottom panel) obtained with the non-local potentials (solid red) and  the  local-equivalent potentials (dotted blue).}\label{Fig1}
\end{figure}

Although transfer cross sections are not usually measured at high energies due to their  low cross sections, it is still interesting to determine the magnitude of the non-local effects for this channel in this energy regime. Indeed, due to the similarities in the probes, one can  expect that  if the effects of non-locality are significant in the transfer channel, they would also be important in stripping, the main contributor to the  knockout cross section~\cite{LB21}. The analysis for transfer is possible thanks to the recent generalization of the adiabatic distorted wave approximation (ADWA) formalism~\cite{JT74} including non-local interactions~\cite{Titus_CPC2016,TitusThesis}. This generalization is implemented in the  {\sc nlat}  code,  available in Ref.~\cite{Titus_CPC2016}.  

To evaluate the magnitude of the non-local effects at high energies, we take as case study the one-neutron transfer reactions $\,^{208}{\rm Pb}(d,p)\,^{209}{\rm Pb}$ at 100~MeV, 138~MeV and 300 MeV. 
In the ADWA formalism, the $d$-$^{208}{\rm Pb}$ adiabatic potential is built from both $n$-$^{208}{\rm Pb}$ and $p$-$^{208}{\rm Pb}$ interactions~\cite{JT74},  evaluated at half the deuteron energy. In addition, the $p$-$^{209}{\rm Pb}$ optical potential in the exit channel is also needed. As in Refs.~\cite{Titus_prc2014,Titus_PRC2016}, we consider the   Perey-Buck non-local potentials~\cite{perey1962}  and  we fit  with {\sc sfresco}~\cite{fresco}   local-equivalent potentials  to the non-local elastic-scattering observables. The parameters for these interactions can  be found in Appendix~\ref{AppOptPot}.  Note that using  local-equivalent potentials for the nucleon-target interactions does not guarantee that  local and non-local adiabatic deuteron potentials are phase-shift equivalent~\cite{TitusThesis,Titus_CPC2016}. 

{As opposed to  \cite{Titus_prc2014,Titus_PRC2016}, in this work we do not  consider non-locality in the bound state wave function. It is well understood that non-locality in the mean field that binds the neutron in the final state  decreases the bound-state wave function in its interior and increases its asymptotic part. Keeping in mind that our aim is to study the effects of non-locality in eikonal models, and the non-local interaction in the bound state calculation is easy to incorporate in these models, here we only focus on the effects of non-locality in the scattering.}

Fig.~\ref{Fig1} shows the transfer cross sections at 100 MeV (top panel), 138 MeV (mid panel) and  300 MeV (bottom panel) obtained  with  non-local (solid red lines) and local-equivalent (dotted blue lines) $n$-$^{208}{\rm Pb}$, $p$-$^{208}{\rm Pb}$ and $p$-$^{209}{\rm Pb}$   potentials.  At all energies, there is a significant effect of non-locality at forward angles. At 100 MeV, non-locality increases the magnitude of the cross section, as already observed in other studies done at lower energy \cite{Titus_PRC2016}, while at larger energies, non-locality reduces it. Similarly, the integrated cross section increases by 22\% at 100~MeV and diminishes by 4\% at 138 MeV and 20\% at 300 MeV. Surprisingly,  non-local effects are rather small at 138~MeV compared to  the ones observed at 100 MeV and 300 MeV.

Consistent with the analysis of transfer reactions at 50~MeV by Titus \etal~\cite{Titus_PRC2016}, we find that non-locality in the deuteron channel has the most influence. Titus \etal  ~explain that this is a result of the combination of two effects: the reduction of the  amplitude of the deuteron scattering wave function in the interior and an additional phase shift coming from the adiabatic description. Because the present work focuses on reactions at higher energies, where the adiabatic approximation is expected to be more accurate,  the deuteron adiabatic local and non-local potentials lead to similar phase shifts.

The fact that  non-locality affects the cross section  differently  at 100 MeV and 300 MeV and does not influence much the transfer observables 138 MeV    can be explained  by the position of the nodes of the incoming and outgoing scattering wave functions. These nodes  cause a compensation of the positive and negative non-local contributions to the $T$-matrix and therefore determine if the non-locality  increases or decreases the cross sections at forward angles. 

Given the large non-local effects on transfer observables at high energies, it is important to also study how non-locality influences knockout  and breakup cross sections in this energy regime. Because eikonal models are the preferred tools to interpret reactions at these energies  we must extend the eikonal approximation to include non-local interactions. 

The development of a non-local eikonal model will be presented in Sec.~\ref{Sec4}. 
However, because when using an eikonal model the effects of non-locality on the reaction observable will be mixed with the effects of the eikonal approximation itself, it is crucial to first establish the level of accuracy that can be expected from this approximation at these energies. This is done in the next section.


\section{Considering approximations to elastic scattering}\label{Sec4}

We consider here the elastic scattering of neutrons on $^{208}\rm Pb$ at 69 MeV and 150 MeV (these energies correspond to half the energy of the cases studied in Sec.~\ref{Sec3}).  The $n$-$^{208}\rm Pb$ interaction  is simulated by the same potentials as the ones used in the previous section, which are detailed in Appendix~\ref{AppOptPot}. 
For completeness, Fig.~\ref{Fig2}  shows the scattering wave function resulting from a local (dotted blue lines) and non-local (solid red lines) interaction. By construction, both  potentials lead to identical wave functions at large distances  and therefore for an exact calculation we expect identical elastic cross sections. What we need to assess is whether this holds under the eikonal approximation.

\begin{figure}
	\centering
	{\includegraphics[clip,trim=0cm 0.5cm 0cm 0cm,width=0.97\linewidth]{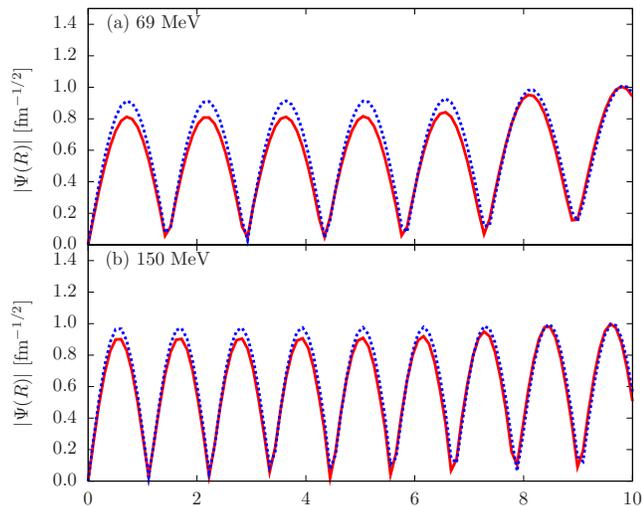}}
	\caption{Scattering wave function for $s$-wave neutron impinging on $^{208}\rm Pb$ at 69~MeV (top panel) and 150~MeV~(bottom panel), obtained with the non-local potentials (solid red) and  the  local-equivalent potentials (dotted blue). }\label{Fig2}
\end{figure}

\begin{figure}
	\centering
	{\includegraphics[clip,trim=0cm 0.5cm 0cm 0cm,width=0.98\linewidth]{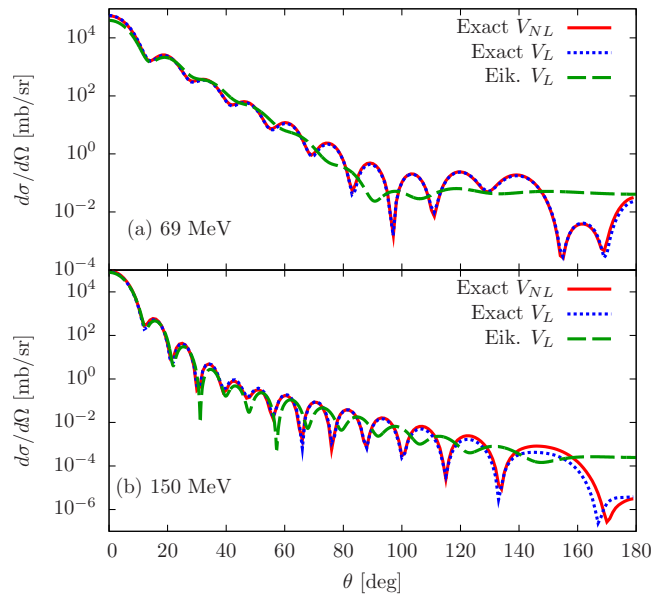}}	
	\caption{Elastic-scattering cross section as a function of the scattering angle $\theta$ for  $^{208}\rm Pb$$(n,n)$$^{208}\rm Pb$ at 69~MeV (top panel) and 150~MeV~(bottom panel): comparison of the results using the exact non-local interaction (red solid line), the exact local-equivalent (blue dotted line) and the eikonal local-equivalent (green dashed line). }\label{Fig3}
\end{figure}

Fig.~\ref{Fig3} shows the elastic-scattering cross section  as a function of the scattering angle  at 69~MeV (top panel) and 150~MeV (bottom panel). The solid red lines are the exact solutions obtained with the non-local potential and the dotted blue lines with the local-equivalent interactions. They agree perfectly with each other except for the largest scattering angles (a limit in the precision of the calculation). Furthermore, the relative difference between the non-local and local-equivalent  integrated elastic-scattering cross sections calculated using exact methods is less than 3\% at 69~MeV and 10\% at 150~MeV~{\footnote{{For the local-equivalent potential at 150 MeV, we have also considered the  prescription by Perey and Buck given in Eq. (35) of Ref.~\cite{perey1962}. It reproduces the non-local cross sections at forward angles but is less accurate than the local-equivalent potential fitted by {\sc sfresco} at larger angles. The corresponding integrated cross section underestimates the non-local one by 17\%.}}}. 
As expected, the reduction of the amplitude of  the scattering wave function in its interior seen in Fig.~\ref{Fig2} does not manifest in the elastic-scattering cross sections which only depends on the asymptotic form of these wave functions.

We now turn to the calculations using the eikonal approximation (dashed green lines in Fig.~\ref{Fig3}).
As expected, the eikonal approximation fails to  describe  the oscillations at large angles at both energies. 
Surprisingly, it also does not reproduce well the forward angles at 69~MeV, for which it underestimates the exact cross section  by 30\% at $0^\circ$. This discrepancy at forward angles is also visible in  integrated elastic-scattering cross sections which are under-predicted by the eikonal approximation by about 21\%.  In constrast, at 150 MeV, the eikonal prediction is accurate up to 50$^\circ$ and reproduces roughly the magnitude of the exact distribution in the whole angular range. At this high energy, the relative difference between the exact and eikonal integrated cross sections obtained with the local-equivalent potential is negligible, i.e.,  about~2\%. 
From this analysis we conclude that the eikonal approximation is  valid for the elastic scattering of neutrons around $150$ MeV \footnote{We should note that for charged particle collisions the situation is different. Then the forward-angles scattering is dominated by the Coulomb interaction, which is typically treated exactly for two-body collisions within the eikonal approximation~\cite{BD04}.  In such a case the eikonal model  would lead  to accurate cross sections for the elastic scattering of charged nuclei, even at energies as low as 50~MeV/nucleon~\cite{AKTB97,HC17,HC18}.}.


\section{Exploring non-local extensions of the eikonal model}\label{Sec5}

\begin{figure}
	\centering
	{	\includegraphics[width=\linewidth]{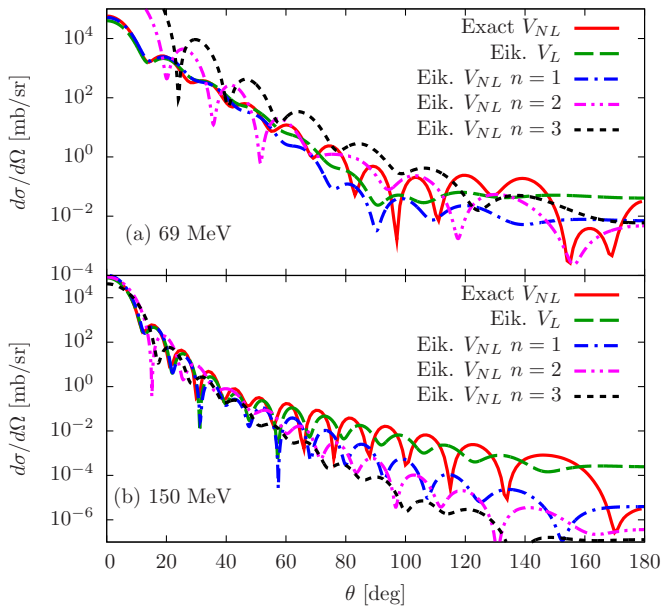}}

	\caption{Elastic-scattering cross section as a function of the scattering angle $\theta$  for  $^{208}\rm Pb$$(n,n)$$^{208}\rm Pb$ at 69~MeV (top panel) and 150~MeV~(bottom panel):
		comparison of the exact non-local (solid red), the eikonal local-equivalent (dashed green) and the non-local eikonal solutions. }\label{Fig4}
\end{figure}

In this section, we study  various extensions of  the eikonal model to include non-local interactions. We  consider the same reaction as before,  the elastic scattering of neutrons on $^{208}\rm Pb$ at $69$~MeV and  $150$~MeV.

As detailed in Sec.~\ref{Sec2}, when the interaction is non-local, the system is described by the non-local \Sch equation~\eqref{eq6}. By reasoning similarly as in the local case and using the same eikonal simplification~\eqref{eq3}--\eqref{eq4}, Eq.~\eqref{eq6}  simplifies into
\begin{equation}
i\hbar v\frac{\partial \widehat \Psi}{\partial Z}(\ve{R})= e^{-iKZ}\int d \ve{R'}\,V_{NL}(\ve{R},\ve{R'}) \widehat\Psi(\ve{R'}) e^{iKZ'}.\label{eq7}
\end{equation}
This equation has formal solutions, which behave asymptotically as
\begin{eqnarray}
\lefteqn{ \widehat \Psi(\ve{R})\underset{Z\to+\infty}{\longrightarrow}}\nonumber\\&&\hspace{-0.8cm}-\frac{i}{\hbar v} \int_{-\infty}^{+\infty}dZ\,\int d \ve{R'}\,V_{NL}(\ve{R},\ve{R'}) \widehat\Psi(\ve{R'}) e^{-iK(Z-Z')}.\label{eq8}
\end{eqnarray}	
Following the idea of Titus \etal~who use an iterative method to include non-local interactions in the ADWA~\cite{TitusThesis,Titus_PRC2016,Titus_CPC2016}, 	 we solve Eq.~\eqref{eq7} iteratively and we take as initial  wave function, the eikonal solution obtained with the local-equivalent potential. 

Fig.~\ref{Fig4} displays the elastic-scattering cross section  at 69 MeV (top panel) and 150 MeV (bottom panel). The exact non-local solution corresponds to the solid red lines, the eikonal local-equivalent to the dashed green lines, and the non-local eikonal at the first, second and third iterations to the dash-dotted blue, the dash-dotted-dotted magenta and the dotted black lines, respectively. At~69~MeV, the first iteration improves slightly the eikonal cross section at $0^\circ$ by increasing its magnitude,   but is less accurate at larger angles. Unfortunately, each additional iteration worsens these results: the cross section is overestimated by several orders of magnitude and the oscillations are not well reproduced. 
Calculations at 150 MeV, for which the eikonal approximation is more accurate, display a similar behavior, exhibiting a slower divergence with the number of iterations.

This failure of the iterative process can be understood by the fact that the non-local potential term  [right hand side of Eq.~\eqref{eq7}] integrates the wave function at the previous iteration over the whole radial space. As detailed in Sec.~\ref{Sec2} and illustrated in Sec.~\ref{Sec4}, the eikonal approximation  is valid only for forward angles. In a semiclassical view, this can be interpreted as the  eikonal description being accurate at large impact parameters,  while it fails at small impact parameters.
Since the non-local term  integrates the eikonal wave function  at the previous iteration  over the whole radial space,  it  accumulates errors at each iteration and the accuracy of the corresponding non-local eikonal cross sections  becomes worst. Naturally, the divergent behavior of the non-local eikonal solution is slower at 150 MeV than 69 MeV since the error made by the eikonal approximation is smaller.  {We have verified that the  scattering amplitude is strongly modified at the first iteration at small $b$s, in the range where the eikonal model describes  poorly the wave function. }{Since the scattering amplitude at the next iteration integrates the wave function at the previous iteration over full space, it is  strongly modified at all impact parameters, causing the  divergence in the elastic-scattering cross section.}

We have investigated two additional implementations of the non-local eikonal solution.  The first one is also an iterative method but now considering that the potential has non-locality in the transverse distance $b$. Although this  non-locality in $b$ is formally easier to handle, ultimately we obtain an expression close to Eq.~\eqref{eq7} in which the non-local potential term still integrates the wave function at the previous iteration over the full space. In this alternative implementation, the wave function  also  accumulates errors at each iteration and  exhibits the same failure of the other iterative process. 

Our third implementation is based on a perturbative eikonal solution to the non-local problem,  detailed in Appendix~\ref{AppA}. Unfortunately, it too leads to an integral  over the full radial range, and eventually diverges with the inclusion of higher orders.

We note that all non-local extensions of the eikonal approximation considered in this work fail for essentially the same reason: the non-local correction is strongly dependent on the short-range description of the scattering wave function, which is not well described by the eikonal approximation. This leads to our conclusion that {in the framework we considered}, the eikonal method is not suitable to handle non-local interactions. Only methods that are able to provide accurate scattering wave functions over the whole radial range can be a useful starting point for   extensions to non-local optical potentials  {based on an iterative approach}.


\section{Conclusions}\label{Conclusions}

Even though in their most general form, optical potentials are non-local, many reaction models have not been adapted to deal with this non-locality.   It has been shown that  non-locality affects strongly transfer observables~\cite{Titus_prc2014,Ross_prc2015,Timofeyuk_prl2013,Waldecker_prc2016,Titus_PRC2016,Ross_prc2016,prc-li2018}. Since these studies were limited to low energy, in this work we  investigated the importance of these effects in the higher energy regime, the regime for which knockout reactions are typically measured.
We  extended the study of Ref.~\cite{Titus_PRC2016} to these energies, i.e., above~50~MeV/nucleon, and analyze the non-local effects  in the projectile-target potentials for $(d,p)$ transfer reactions. Our results show that non-locality affects strongly transfer angular distributions and therefore are likely to influence significantly the stripping process, which is the largest contributor to  knockout observables.

Because knockout reactions are usually analyzed with eikonal models, we then investigated the extension of this theory to include non-local interactions. We considered  elastic scattering of neutrons on a $^{208}\rm Pb$ target.  Following the same idea used  in 
Ref.~\cite{TitusThesis,Titus_CPC2016}, we adopted an iterative method to obtain the solution of the non-local scattering equation, in which  the non-local  correction involved a radial integration of the product of eikonal wave function at the previous iteration and the non-local potential.  Our results show that this iterative solution diverges because the eikonal solution is not accurate at short distances,  causing an  accumulation of errors at each iteration. Other approaches were considered but turned out to suffer from the same problem.

This analysis suggests that models that provide an accurate description of the wave function over its whole radial range are better suited to describe high-energy reactions such as knockout and breakup, when incorporating non-local interactions {iteratively}. This includes the distorted wave Born approximation, the continuum-discretized coupled channel method~\cite{Kam86,YOMM12} or the dynamical eikonal approximation~\cite{BCG05}.


\begin{acknowledgments}
The authors would like to thank  P. Capel for  useful discussions. The work of C. H.   is supported by the U.S. Department of Energy, Office of Science, Office of Nuclear Physics, under the FRIB Theory Alliance award no. DE-SC0013617 and {under Work Proposal no. SCW0498}. This work was prepared in part by LLNL under Contract no. DE-AC52-07NA27344.
This work was supported by the National Science Foundation under Grant  PHY-1811815 and the U.S. Department of Energy grant DE-SC0021422.  This work relied on iCER and the High Performance Computing Center at Michigan State University for computational resources. 
\end{acknowledgments}

\appendix\section{Choice of  potentials}
\label{AppOptPot}
\begin{table*}
	\centering
	\begin{tabular}{ccc|ccccccccccc}
		&&&$V_R$&$R_R$&$a_R$&$W_I$&$R_I$&$a_I$&$W_D$&$R_D$&$a_D$&$\beta$&$\chi^2/N$\\
		&&&[MeV]&[fm]&[fm]&[MeV]&[fm]&[fm]&[MeV]&[fm]&[fm]&[fm]&\\\hline\hline
		\multirow{3}{*}{$V_{NL}$}&$n$-$^{208}\rm Pb$&&\multirow{2}{*}{71}&\multirow{2}{*}{7.229}&\multirow{2}{*}{0.650}&&&&\multirow{2}{*}{15}&\multirow{2}{*}{7.229}&\multirow{2}{*}{0.470}&\multirow{2}{*}{0.85}&\multirow{2}{*}{}\\
		&$p$-$^{208}\rm Pb$&&&&&&&&&&&&\\
		&$p$-$^{209}\rm Pb$&&71&7.240&0.650&&&&15&7.240&0.470&0.85&\\\hline
		\multirow{9}{*}{$V_{L}$}&\multirow{3}{*}{$n$-$^{208}\rm Pb$}&50 MeV&34.153&7.435&0.610&&&&8.174&7.314&0.422&&$\chi^2/N =0.229$\\
		&&69 MeV&29.475   &7.416& 0.621&0.280   &8.876 &0.400&6.711  &7.318&    0.400&& $\chi^2/N = 2.237$\\
		&&150 MeV&15.783   &7.355    &0.580&&&&4.005&
		7.284 &   0.406&&$\chi^2/N =6.600$\\\cline{2-14}
		&\multirow{3}{*}{$p$-$^{208}\rm Pb$}&50 MeV&38.969&7.434&0.615&&&&9.030&7.324&0.420&&$\chi^2/N =0.261$\\
		&&69 MeV&33.755   &7.424&0.606&&&&7.818    &7.332&0.415&&$\chi^2/N = 1.282$\\
		&&150 MeV& 20.303  &  7.244   & 0.639&&&& 4.672&
		7.274    &0.401&&$\chi^2/N =4.234$\\\cline{2-14}
		&\multirow{3}{*}{$p$-$^{209}\rm Pb$}&101.7 MeV&26.092&7.412&0.612&&&&5.867&7.339&0.398&& $\chi^2/N =2.401$\\
		&&139.7 MeV& 17.790&    7.426   &0.596&&&&3.857&
		7.355    &0.369&&$\chi^2/N = 0.895$\\			
		&&301.7 MeV& 6.151&    7.029&    0.614&&&&1.434  & 7.276&    0.400&&$\chi^2/N =3.365$\\			
	\end{tabular}
	\caption{Parameters of the  potentials: the non-local interactions are taken from Ref.~\cite{perey1962} and their local-equivalent are fitted with {\sc sfresco}~\cite{fresco}. }\label{TabOptPot}
\end{table*}

In this paper, we study the one-neutron transfer reaction $\,^{208}{\rm Pb}(d,p)^{209}{\rm Pb}$ at 100~MeV, $138$~MeV and 300~MeV. To model these reactions within the ADWA, we use  single-particle local potentials to generate the  bound states, and optical potentials to simulate the projectile-target interactions in the entrance and exit channels.

 As in Ref.~\cite{Titus_CPC2016}, the deuteron bound state is produced  by a Gaussian potential of range 1.494 fm and a depth of 71.85 MeV. We also adopt the same description of  $^{209}{\rm Pb}$ as in Ref.~\cite{Titus_CPC2016}, using a real single-particle $^{208}\rm Pb$-$n$   potential composed of a volume and a spin-orbit term. The central Woods-Saxon shape is characterized by a radius of 7.406 fm and a diffuseness of 0.65 fm. The depths are fitted to reproduce the valence neutron binding energy of $^{209}\rm Pb$, the real depth is given by 46.561  MeV and the spin-orbit strength by 6~MeV.
 
The optical potentials  needed for the ADWA reaction model are $V(n$-$^{208}{\rm Pb})$, $V(p$-$^{208}{\rm Pb})$ at half the deuteron incident energy  and $V(p$-$^{209}{\rm Pb})$ at the energy in the exit channel.
We take the non-local interaction developed by Perey and Buck~\cite{perey1962} defined by
\begin{eqnarray}
V_{NL} (\ve R,\ve R') &=&V(\tilde R) \frac{\exp \left[-\left(\frac{| \ve R- \ve R'|}{\beta}\right)^2\right]}{\pi^{3/2}\beta^3}.\label{eq9}
\end{eqnarray}
with $\tilde R=(R+R')/2$.
The local part of this potential is parametrized with a Woods-Saxon form as
\begin{eqnarray}
V (R) &=& -V_R\,f_{\rm WS}(R,R_R,a_R) -i\,W_I\,f_{\rm WS}(R,R_I,a_I) \nonumber \\
&& +i\,4 a_D W_D \deriv{}{R} f_{\rm WS}(R,R_D,a_D),\label{eq10}
\end{eqnarray}
where
\begin{equation}
f_{\rm WS}(R,R_X,a_X) = \frac{1}{1 +e^{\frac{R-R_X}{a_X}}}.\label{eq11}
\end{equation}
The parameters of the Perey-Buck interactions are given in Table~\ref{TabOptPot}~\footnote{We neglect the spin-orbit term for simplicity.}. This potential is energy-independent, therefore the same parameters are  used for all energies considered.

For a meaningful comparison, we build the local-equivalent potentials $V_L$ by fitting with {\sc sfresco}~\cite{fresco} the exact elastic-scattering observables obtained  from the non-local interactions and with an artificial relative error of 10\%. These potentials are parametrized with a Woods-Saxon form~\eqref{eq10}--\eqref{eq11} and the corresponding parameters  are displayed in Table~\ref{TabOptPot}.  In the last column, we present  the $\chi^2/N$ resulting from each fit.

Of course, the Coulomb interaction is local and we take it to be that of a uniformly charged sphere of radius $R_C=1.25\times A_T^{1/3}$~fm, with $A_T$ the mass number of the target.

\section{Perturbative non-local eikonal solutions}\label{AppA}

In this appendix, we study an alternative eikonal solution to the non-local \Sch equation~\eqref{eq6}. Following a perturbative approach, we write   the non-local potential as a sum of a local and non-local terms, i.e, $V_{NL}(\ve R,\ve R')=V_L( R)+\Delta V_{NL}(\ve R,\ve R')$, and  treat $\Delta V_{NL}$  as a perturbation.  We take for  $V_L$ the local-equivalent potential and for $\Delta V_{NL}=V_{NL}-V_L $   the difference between the non-local potential and the local-equivalent one.

Accordingly, the eikonal wave functions can  be expressed as the sum of a leading term $\widehat\Psi^0$ and a perturbation $\widehat\Psi^1$
\begin{eqnarray}
\widehat {\Psi}(\ve R)=\widehat\Psi^0(\ve R)+\widehat\Psi^1(\ve R).
\end{eqnarray}
In the eikonal approximation, the leading term $\widehat\Psi^0$ is simply the local-equivalent eikonal solution~\eqref{eq5}.
The first-order term is obtained from the non-local eikonal equation
\begin{eqnarray}
i\hbar v\frac{\partial\widehat\Psi^1}{\partial Z}(\ve R)&=&V_{L}(R)\widehat\Psi^1(\ve R)\nonumber \\
&&\hspace{-1.8cm}+e^{-iKZ}\int d\ve{R'} \, \Delta V_{NL}(\ve{R},\ve{R'}) e^{iKZ'}\widehat\Psi^0(\ve{R'}).\label{eq13}
\end{eqnarray}
Note that the non-local term of this equation depends on the eikonal solution  $\widehat\Psi^0$, similarly as in the  iterative method explored in Sec.~\ref{Sec4}.
The non-local eikonal perturbative equation~\eqref{eq13} can be solved analytically, its solutions tend  to
\begin{eqnarray}
&&\widehat\Psi^1(\ve R)\underset{Z\to+\infty}{\longrightarrow}-\frac{i}{{\hbar v}}
e^{-\frac{i}{\hbar v}\int_{-\infty}^{+\infty} dZ\,V_{L}(R)}\nonumber\\
&&\times \quad \int_{-\infty}^{+\infty} dZ\, e^{-iKZ}e^{\frac{i}{\hbar v}\int^{Z}_{-\infty}dZ'\,V_{L}(R')}\nonumber\\
&&\times \quad\int d\ve{R'} \,\Delta V_{NL}(\ve{R},\ve{R'}) e^{iKZ'}\widehat\Psi^0(\ve{R'}).
\end{eqnarray}

\begin{figure}
	{	\includegraphics[width=\linewidth]{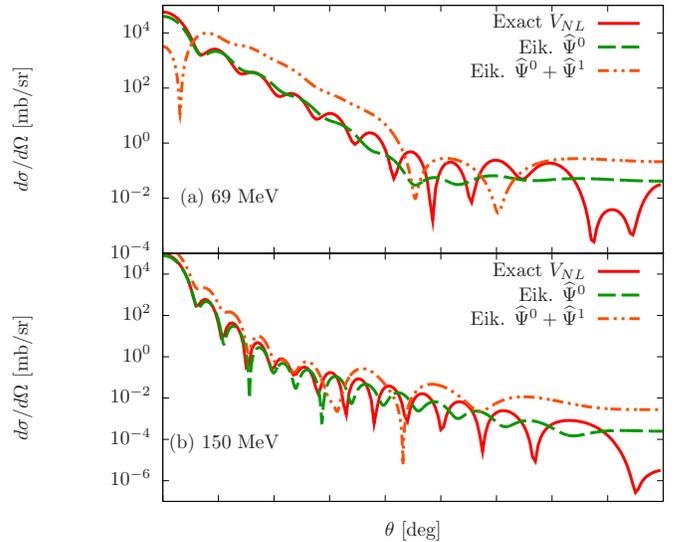}}
	
	\caption{Elastic-scattering cross section  as a function of the scattering angle $\theta$ for $^{208}\rm Pb$$(n,n)$$^{208}\rm Pb$ at 69~MeV (top panel) and 150 MeV (bottom panel): comparison of the exact non-local (solid red) and the perturbative eikonal solutions.}\label{FigA1}
\end{figure}

Fig.~\ref{FigA1} shows  the elastic-scattering cross section for neutron scattering on $^{208}\rm Pb$ target at 69~MeV (top panel) and 300 MeV (bottom panel), as a function of the scattering angle. 
The leading-order eikonal perturbative solution (dashed green lines) is simply the local-equivalent eikonal cross section. As already noted in Sec.~\ref{Sec4}, it  reproduces  well the magnitude of the exact non-local cross sections (solid red lines) at $0^\circ$ at $300$~MeV energies but not at 69~MeV. Also, it  is not accurate at larger angles, {mostly due to the eikonal approximation which is inadequate at the largest angles}. 

The cross sections obtained with the first-order perturbation of the non-local eikonal solution~\eqref{eq11} are plotted by the dash-dotted-dotted orange lines. At both energies, the first-order calculation worsens the result. Even when considering a different leading order solution, viz. by using the local diagonal part of the non-local potential for $V_L$, we arrive at the same issue. 
	 Just as in the analysis of the iterative solution in Sec.~\ref{Sec4}, this failure can be explained by the inaccuracy of the eikonal wave function at short distances which leads to large errors in the integral associated with the non-local correction. Unfortunately, this perturbative approach   also fails to extend the eikonal method to include non-local interactions.

\end{document}